\documentclass[twocolumn]{aastex631}
\usepackage{graphicx}
\usepackage{dcolumn}
\usepackage{bm}
\usepackage[T1]{fontenc}
\usepackage{mathptmx}
\usepackage{mathrsfs}
\usepackage{xcolor}
\usepackage{hyperref}
\usepackage{physics}
\usepackage{simplewick}
\usepackage{mathrsfs}
\usepackage[bottom]{footmisc}
\usepackage{threeparttable}
\usepackage{appendix}

\newcommand{\msun}{M_{\odot}}

\shorttitle{Model for TDE and QPE in GSN 069}
\shortauthors{Wang et al.}
\graphicspath{{./}{figures/}}
\newcommand{\R}[1]{\textcolor{red}{#1}}
\newcommand{\B}[1]{\textcolor{blue}{#1}}

\begin{document}
\title{A model for the possible connection between tidal disruption event and quasi-periodic eruption in GSN 069} 
\author[0000-0001-5019-4729]{Mengye Wang}
\affiliation{Department of Astronomy, School of physics, Huazhong University of Science and Technology, Luoyu Road 1037, Wuhan, China}
\author[0000-0002-2670-2305]{Jinjing Yin}
\affiliation{Department of Astronomy, School of physics, Huazhong University of Science and Technology, Luoyu Road 1037, Wuhan, China}
\author[0000-0001-7192-4874]{Yiqiu Ma$^*$}
\email{* Corresponding author: myqphy@hust.edu.cn}
\affiliation{Department of Astronomy, School of physics, Huazhong University of Science and Technology, Luoyu Road 1037, Wuhan, China}
\affiliation{Center for Gravitational Experiment, School of physics, Huazhong University of Science and Technology, Luoyu Road 1037, Wuhan, China}
\author[0000-0003-4773-4987]{Qingwen Wu$^*$}
\email{* Corresponding author: qwwu@hust.edu.cn} 
\affiliation{Department of Astronomy, School of physics, Huazhong University of Science and Technology, Luoyu Road 1037, Wuhan, China}

\begin{abstract}
Quasi-periodic eruptions (QPEs) are found in the center of five galaxies, where a tidal disruption event (TDE)-like event is also reported in GSN 069 that happened a couple of years before the QPEs. We explain the connection of these phenomena based on a model of a highly eccentric white-dwarf (WD)-$10^{4-6}M_{\odot}$ massive black hole (MBH) binary formed by the Hill mechanism. In this system, the tidal induced internal oscillation of WD can heat the WD envelope thereby induces the tidal nova and inflates the WD envelope, which can be captured by the MBH and form a TDE. The tidal stripping of the surviving WD in the eccentric orbit can produce the QPEs. We also apply this model to the other four QPE sources. Based on the estimated fallback rate, we find that the remaining time after the QPE-observed time for these QPEs is only around 1-2 years based on our simple model estimation, and then the WD will be fully disrupted. We also show that the accretion rate can be much higher than Eddington accretion rate in final stage of these QPE sources. The peak frequency of spectral energy distribution of disk stays in soft X-ray band ($\sim 0.1-1$ keV), which is consistent with observational results. 
\end{abstract}


\section{Introduction}    \label{sec:intro}

Recently discovered regular X-ray bursts are reported as quasi-periodic eruptions (QPEs) in five galaxies, which recur from 2.4-hour (eRO-QPE2) to 18.5-hour (eRO-QPE1) timescale with duration varies from 0.5 hour (eRO-QPE2) to 7.6 hours (eRO-QPE1)\,\citep{Miniutti2019,Giustini2020,Arcodia2021,Joheen2021}. The first QPE source GSN 069 detected by XMM-Newton in December 24, 2018, has a peak X-ray luminosity of $L\sim 5\times 10^{42}$\,erg/s and a nine-hour recurrence time\,\citep{Miniutti2019}. The second QPE source is RX J1301.9+2747 with an average 5-hour separation between X-ray peak flux, which was found in the XMM-Newton observations of 2019 and in the archival observational data in 2000\,\citep[][]{Giustini2020}. It should be noted that RX J1301.9+2747 has no very strict recurrence time since that the first two peaks are separated by a longer recurrence time (about 20 ks) compared to the second and third peaks (about 13 ks)\,\citep[see][]{Giustini2020}. The other QPE events eRO-QPE1/2 with a 18.5/2.4 hours recurrence time were discovered in two low-luminosity active galaxies by eROSITA\,\citep{Arcodia2021}. In addition, the fifth QPE candidate XMMSL1 J024916.6-041244 shows two QPE-like flares in soft X-rays in 2006\,\citep{Joheen2021}. These QPEs have a coincidence with the galactic nuclei, where all host galaxies are found to be \emph{dwarf galaxies} with total stellar mass around $m_\star \sim 10^9M_\odot$\,\citep{Arcodia2021} and black hole (BH) mass of $\sim10^{5-6}\msun$\,\citep[e.g.,][]{Wevers2022}. Interestingly, for GSN 069, historical data shows that it was a quiescent galaxy in 1989, while its X-ray emission is brightened by a factor of at least 240 in 2010\,\citep{Miniutti2013,shu2018}. However, the QPEs were not detected during the initial brightening stage. Naturally it is interesting to conjecture that this initial brightening and the later observed QPEs may be attributed to the connected physical processes in the same astrophysical source.

Currently, several models have been proposed to explain the physical source/mechanism of the QPE events. For example, radiation pressure instabilities in the standard thin accretion disk at high accretion state\,\citep{Janiuk2002,Miniutti2019,Arcodia2021} was predicted to trigger these QPEs. However, there exists inconsistency between the model predictions and the observational data\,\citep[][]{Arcodia2021}. \cite{Pan2022} suggested that the radiation-pressure instability model can explain the QPEs if a non-zero viscous torque condition is included on the inner boundary of disk. This model can explain the QPEs in active galaxies, but is difficult to explain the QPEs that observed in low-luminosity AGNs\,\citep[e.g., eRO-QPE1/2,][]{Wevers2022}. \cite{Xian2021} proposed that QPE of GSN 069 can be driven by star-disk collision, while this model can explain the QPEs but cannot explain its initial brightening before the QPEs. Alternatively, partial disruptions that happen in an extreme-mass-ratio inspiral\,(EMRI) system, e.g. a solar mass compact object (CO) orbiting around a massive BH (MBH) with mass $10^{4-6}M_{\odot}$\,\citep[see][for a recent review]{kate2020}, are also predicted to drive the QPEs\,\citep[e.g.,][]{King2020,WangFayin2021}. In particular, \citet{King2020} proposed that a white dwarf (WD) EMRI with a highly eccentric orbit can reproduce the nine-hour QPEs as found in GSN 069, where the periodical outflow from the Roche lobe onto the MBH can trigger QPEs\,\citep[see the theoretical work in][]{Zalamea2010}. To form a binary system with a highly eccentric orbit, one typical way is the Hills mechanism. However, \cite{metzger2021} suggested the event rate of QPEs by Hills mechanism may be too low to explain the observation, therefore a new model based on interacting circular stellar EMRI pairs, which is formed through either the tidal separation of binaries (Hills mechanism) or Type I inward migration through a gaseous AGN disk.

In this work, we propose a model to connect the TDE-like flare and the followed QPEs as found in GSN 069 based on a WD binary is tidally detached by a MBH, where the TDE-like event is triggered by the tidal disruption of WD envelope and the QPEs are produced by the tidal disruption of the surviving WD. The other four QPEs are also discussed based on the above model. We present the model and physical processes in Section\,\ref{sec:model}, main results and discussions are presented in Section\,\ref{sec:results and discussions}. In Section\,\ref{sec:summary}, we summarize our main results.

\section{Model and physical processes} \label{sec:model}

\begin{figure}
\centering
\includegraphics[scale=0.5]{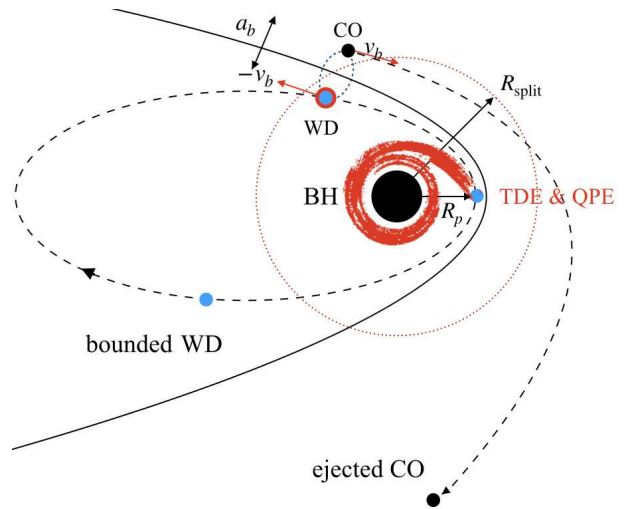}
\caption{Schematic picture of our model. A WD binary is captured and separated by the tidal field of a MBH. The WD is bounded to the MBH and form a highly eccentric WD-MBH EMRI system, while the other compact object (CO) is ejected away. A tidal nova explodes the WD envelope, which is then captured by the MBH and thereby forms a TDE. A couple of years later, the tidal stripping of the remaining WD starts and produces the QPEs.}
\label{fig:1}
\end{figure}

To produce the QPEs with a period of several hours, a highly eccentric orbit of WD-MBH system is required\,\citep[e.g.,][]{King2020}. The Hills mechanism is one typical formation channel of such a highly eccentric orbit shown schematically in Figure\,\ref{fig:1}, where the tidal separation of binaries through a 3-body exchange between a WD+CO (e.g., white dwarf, neutron star or stellar-mass BH) binary and a MBH\,\citep[e.g.,][]{Hills1988,Bromley2012,Brown2015}. When the WD binary is approaching a MBH following a parabolic orbit, the tidal force can split the binary. The splitting happens at an orbit radius $R_{\rm split}\sim a_b (M_{\rm BH}/M_{b})^{1/3}$, where the $M_{\rm BH}$ is the mass of the MBH, $a_b$ and $M_b$ is the separation and total mass of the WD binary, respectively \,\citep[e.g.,][]{Pau2020}.  One binary component is bounded to the MBH, while the other will be ejected and form a hypervelocity star\,\citep[]{Hills1988}. If the WD is captured, the semi-major axis of the new bound orbit is $a_{\rm cap}\sim a_b(M_{\rm BH}/M_b)^{2/3}$\,\citep{Pau2018,Pau2020}, where $M_{\rm WD}$ and $R_{\rm WD}$ are mass and radius of WD, respectively. The triggering of the WD tidal stripping requires that the pericenter distance of the bound WD satisfies $R_p=(1-e_0)a_{\rm cap}\sim 2R_t$\,\citep{Zalamea2010}, where $R_t=R_{\rm WD}(M_{\rm BH}/M_{\rm WD})^{1/3}$ is the tidal radius and $e_0$ is the initial eccentricity. The initial eccentricity of the bound WD can be estimated with $e_0\sim1-(R_{p}/R_{\rm split})(M_{\rm BH}/M_b)^{-1/3}$\,\citep[][]{Pau2018,Pau2020}, which is shown in Appendix \ref{appendix}. For $M_{\rm BH}\sim 10^5M_\odot$, $M_b\sim M_\odot$ ($0.5+0.5M_\odot$ binary) and $a_b\sim (5-50)R_{\rm WD}$, the eccentricity $e_0\sim0.97-1$.

This system could produce both the TDE-like flare and the later QPEs phenomenon as follows.

(I) The WD will be tidally dynamically deformed by the MBH before the tidal disruption. During the passage of the pericenter, the internal oscillation modes of the WD can be excited by the time-varying tidal force, which leads to the exchange of energy and momentum between the orbital motion and the stellar oscillations. If the WD has an envelope, the energy of stellar oscillation could deposit energy and heat the envelope, which is the \emph{tidal heating} process\,\citep[e.g.,][]{fuller2012tidal,vick2017tidal,Yang2018Evolution}. Such a process could induce a runaway fusion in the hydrogen/helium envelope of the WD before its disruption, which is called tidal nova\,\citep{fuller2011tidal,fuller2012dynamical}. The envelope ignited by the heat released in these processes may inflate, and over-fill the Roche lobe thereby create a TDE, which could be an explanation to the X-ray TDE-like flare of GSN 069 in 2010.

(II) After the tidal nova, the surviving WD core is left on the eccentric orbit. The orbit will shrink due to the loss of energy and angular momentum by various physical processes such as gravitational radiation, tidal oscillation and mass transfer, which will be discussed in detail in Section\,\ref{sec:2.2}. A few years after the tidal nova of the envelope, the WD core can over-fill the Roche lobe radius and the tidal stripping will start. In contrast to the full tidal disruption in the tidal nova stage, the tidal stripping occurs at each pericenter passage\,\citep[see also,][]{Zalamea2010,macleod2014illuminating}.

\subsection{A tidally deformed white dwarf}

For an undeformed WD, we adopt the model used in \cite{Zalamea2010}. The WD mass-radius relation is
\begin{equation} \label{1}
    R_{\rm WD}=9.04\times 10^8\left(\frac{M_{\rm WD}}{M_{\rm Ch}}\right)^{-1/3}
\left(1-\frac{M_{\rm WD}}{M_{\rm Ch}}\right)^{0.447} \quad  \rm cm ,
\end{equation}
\noindent where $M_{\rm Ch}=1.44M_\odot$ is the Chandrasekhar mass. To estimate the surface density of the WD, we adopt the polytropic equation of state in the surface layer $P=K\rho^\gamma$, where $\gamma=5/3$, $K\equiv(3/\pi)^{2/3}h^2/20m_e(\mu_e m_p)^{5/3}$ for non-relativistic degenerate electrons with Planck's constant $h$, electron mass $m_e$, proton mass $m_p$ and $\mu_e\approx 2$ is the mean molecular weight per electron. Integrating the hydrostatic equation in the radial direction $dP/dz=\rho GM/R_{\rm WD}^2$ leads to
\begin{equation} \label{2}
    \rho(z)=\left(\frac{2GM_{\rm WD}}{5KR_{\rm WD}}\right)^{3/2}\left(\frac{z}{R_{\rm WD}}\right)^{3/2},
\end{equation}
\noindent where $G$ is gravitational constant, $\rho$ is the density and $z$ is the surface depth of the WD. 

Many WDs are surrounded by a thin envelope of \emph{non-degenerate} helium/hydrogen with a mass in the range of  $10^{-15} $ to $10^{-2}M_\odot$, which depends on the stellar evolution history\,\citep[e.g.,][]{Isern2004}. For a typical mass of envelope with $10^{-4}M_{\rm WD}$, it occupies about 10 percent of the WD's radius\,\citep{Iben1984,romero2019white}.

For a WD-MBH binary system, the WD will be tidally deformed by its companion MBH, which excites the stellar oscillations in the WD\,\citep[mainly g-modes and f-mode, see][]{IP2004,IP2007,vick2017tidal,Yang2018Evolution}. The WD internal-mode frequency depends on the concrete WD models, where the frequency $\omega_g\sim (0.01-0.1)\Omega_\star$ for g-modes and $\omega_f\sim \Omega_\star$ for f-mode\,\citep[$\Omega_\star=\sqrt{GM_{\rm WD}/R_{\rm WD}^3}$e.g.,][]{1973Nonradial,Lee1986,E2006The,fuller2011tidal,vick2017tidal}. The WD stellar-oscillation modes can interact with the envelope through the non-linear wave interaction and thereby deposit its energy into the envelope. The tidal induced energy dissipation rate is very sensitive to the pericenter distance\,\citep[e.g.,][]{IP2004,IP2007,vick2017tidal}. In the case of small $\eta\equiv R_p/R_t \sim 2-3$, the bypass of the WD near the pericenter can quickly trigger the tidal nova, while a sufficiently long time is needed to accumulate the energy for the larger $\eta$. Moreover, this tidal interaction between the orbital motion and the WD internal oscillation can back-act on the orbital evolution process.

\subsection{The binary orbital evolution} \label{sec:2.2}
The eccentric orbit of WD-MBH is characterised by the orbital semi-major axis $a$ and eccentricity $e$. Orbital evolution of $a$ and $e$ is mainly contributed by the gravitational wave (GW) radiation ($\dot{a}_{\rm GW}$), the dynamical tides (DT) to the WD oscillation ($\dot{a}_{\rm DT}$) and the mass transfer (MT) rate ($\dot{a}_{\rm MT})$: $\dot{a}=\dot{a}_{\rm GW}+\dot{a}_{\rm DT}+\dot{a}_{\rm MT}$.

The change of orbital parameters due to the gravitational radiation is given by\,\cite{peters1964}:
\begin{equation} \label{4}
    \dot{a}_{\rm GW}=-\frac{32G^3M_{\rm BH} M_{\rm WD} (M_{\rm BH}+M_{\rm WD}) }{5c^5a^3(1-e^2)^{7/2}}\left(1+\frac{73}{24}e^2+\frac{37}{96}e^4\right),
\end{equation}
\begin{equation} \label{5}
    \dot{e}_{\rm GW}=-\frac{304}{15}e\frac{G^3M_{\rm BH} M_{\rm WD} (M_{\rm BH}+M_{\rm WD}) }{c^5a^4(1-e^2)^{5/2}}\left(1+\frac{121}{304}e^2\right).
\end{equation}

\begin{figure}
\centering
\includegraphics[scale=0.28]{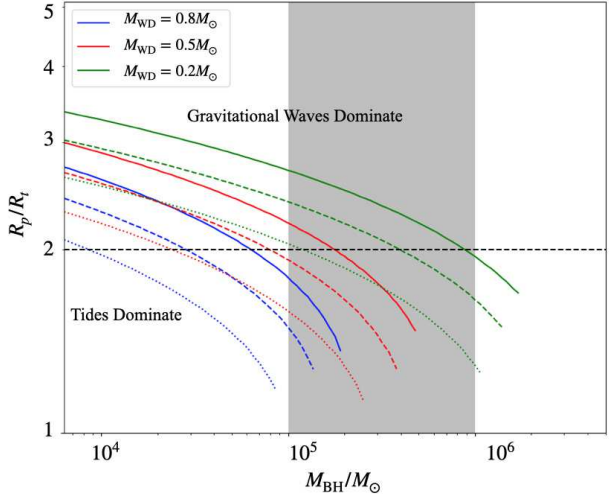}
\caption{Trajectory of WD-MBH binaries evolution in the parameter phase space. The green, red and blue lines represent $M_{\rm WD}=0.8$, $0.5$ and $0.2M_\odot$, respectively, where the solid, dashed and dotted lines represent $\widetilde{\Omega}_s=0$, 0.3 and 0.5, respectively. The black dashed line corresponds to $R_p/R_t=2$. Gravitational radiation 
dominates the orbital evolution above these trajectory lines, while the tidal excitation is the dominant orbital energy loss source when the pericenter distance locates below these trajectory lines. The gray shaded area shows the region of the typical MBH mass for the five interested QPEs sources ($10^{5-6}M_\odot$).}
\label{fig:0}
\end{figure}

Comparing to the angular momentum loss caused by gravitational radiation, the dynamical tides in the g-modes have a negligible impact on the orbital evolution\,\citep[][]{vick2017tidal}. To estimate the orbital energy and angular momentum transfer in one orbit by the dynamical tides in f-mode, we follow the method developed by \cite{IP2004} \citep[see also,][]{Yang2018Evolution}:
\begin{equation}\label{6}
\Delta E_{\rm DT}=\frac{16\sqrt{2}}{15}\widetilde{\omega}_f^3\widetilde{Q}_f^2\eta^{3/2}{\rm exp}[-\frac{4\sqrt{2}}{3}\eta^{3/2}(\widetilde{\omega}_f+\widetilde{\Omega}_s)]\frac{GM_{\rm WD}^2}{R_{\rm WD}},
\end{equation}
where $\widetilde{\Omega}_s\equiv \Omega_s/\Omega_\star$ is the dimensionless rotation angular velocity of the WD, $\widetilde{\omega}_f\equiv \omega_f/\Omega_\star=1.455$ is the dimensionless frequency of the WD $f$-mode and $\widetilde{Q}_f\equiv Q_f/\sqrt{M_{\rm WD}}R_{\rm WD}=0.5$ is the dimensionless overlap of the companion tidal field with the wave-function of the WD internal $f$-mode calculated in\,\cite{Lee1986}. It is important to note that Eq.(\ref{6}) is only valid for $\widetilde{\Omega}_s\le 0.5$\,\citep[][]{IP2007}. With above typical parameters ($\widetilde{\Omega}_s=1.455$ and $\widetilde{Q}_f=0.5$), the Eq.(\ref{6}) can be written as $\Delta E_{\rm DT}=0.07\eta^{3/2}{\rm exp}[-2.74(\eta^{3/2}-1)]\Psi GM_{\rm WD}^2/R_{\rm WD}$, where $\Psi={\rm exp}(-\frac{4\sqrt{2}}{3}\widetilde{\Omega}_s\eta^{3/2})$. \footnote{Note that there is a typo of Eq.(7) in \cite{IP2007}, where 0.7 should be 0.07. Therefore the dynamical tidal effects are overestimated in\,\cite{IP2007} and\,\cite{macleod2014illuminating}}\label{footnote:1}
The orbital angular momentum transfer is related to $\Delta E_{\rm DT}$ by
\begin{equation}\label{eq:LDT}
\Delta L_{\rm DT}=2\Delta E_{\rm DT}/\omega_f.
\end{equation}
The rotation of WD can play an important role in the tidal coupling process\,\citep[see above Eq.\,\ref{6},][]{IP2004,IP2007}. \cite{vick2017tidal} proposed that the rotation has a small effect on dynamical tides for a captured isolated WD, while the rotational effect should be important if the WD is captured through a binary, which happens in our model. In Figure\,\ref{fig:0}, we show the WD-MBH binaries evolution trajectories in the parameter phase space, which is revised from the result given in\,\cite{macleod2014illuminating} based on the overestimated dynamical tidal effect\,(see the footnote below Eq.\,\ref{eq:LDT}). Gravitational radiation is the dominant orbital evolution term above these trajectory lines, while the tidal excitation dominates the orbital energy loss when the pericenter distance evolves below these trajectory lines. For the typical MBH mass of the interested QPEs\,\citep[see][]{Wevers2022}, we find that the orbital angular momentum transfer caused by the dynamical tides are still weaker than that of gravitational radiation, except for the WDs with extremely small mass and very slow rotation rate. We will take into account  this tidal interaction process in our following calculations of orbital evolution without considering the tidal heating effect.

When the Roche lobe shrinks to the surface of the WD, the tidal stripping will start. The mass transfer induced semi-major axis and the eccentricity orbital average change are given by\,\citep{Sepinsky2007b}:
\begin{equation} \label{8}
\dot{a}_{\rm MT}=\frac{a}{\pi}\frac{\dot{M}_{\rm WD}}{M_{\rm WD}}(q-1)(1-e^2)^{1/2},
\end{equation}
\begin{equation} \label{9}
\dot{e}_{\rm MT}=\frac{1}{\pi}\frac{\dot{M}_{\rm WD}}{M_{\rm WD}}(q-1)(1-e^2)^{1/2}(1-e),
\end{equation}

\noindent where $\dot{M}_{\rm WD}$ is the mass transfer rate ($\dot{M}_{\rm WD}<0$) and $q\equiv M_{\rm WD}/M_{\rm BH}$ is the mass ratio. According to the Eq.\,(\ref{4}) and Eq.\,(\ref{8}), we find $\delta a_{\rm MT}/\delta a_{\rm GW} \sim 10^3 \delta M_{\rm WD}/M_{\rm WD} $ for eccentricity $\sim 0.98$ in one orbital period, where $\delta M_{\rm WD}$ is the mass loss of WD. The orbital evolution should be dominated by the mass transfer when $\delta M_{\rm WD}/M_{\rm WD} >10^{-3}$, which will happen in last several tens of orbits before the WD was fully disrupted. 

In summary, we consider the orbital evolution caused by gravitational radiation, mass transfer and tidal effect, where the gravitational wave and tidal effects shrink the orbit while the mass transfer has the opposite effect. Combining the relations of the pericenter radius $R_p=a(1-e)$, orbital energy $E=-GM_{\rm BH}M_{\rm WD}/2a$ and orbital angular momentum $L=\sqrt{GM_{\rm BH}M_{\rm WD}^2a(1-e^2)}$, we obtain the evolution of pericenter radius as:
\begin{equation}\label{Rp}
\dot{R}_p=\sqrt{\frac{(M_{\rm BH}+M_{\rm WD})a(1-e^2)}{GM^2_{\rm BH}M^2_{\rm WD}}}\frac{\dot{E}}{\Omega_p} \left[\frac{\dot{L}\Omega_p}{\dot{E}}-\frac{1}{\sqrt{1+e}}\right],  
\end{equation} 
where $\Omega_p=\sqrt{GM_{\rm BH}/R_p^3}$ is the Kepler frequency at the pericenter. The $\dot{L}/\dot{E}$ is governed by both the gravitational radiation and the tidal-coupling. The value of $\dot{R}_{p}$ is negative and positive for the gravitational-radiation dominated case and the dynamical tidal-coupling dominated case, respectively.

\subsection{Tidal disruption and tidal stripping of WD } \label{subsec:2.3}

When the WD passes close to the MBH at pericenter before tidal stripping starts, the tidal force from the central MBH excites internal waves in the WD, transferring energy and angular momentum between the WD and its orbit\,\citep{fuller2011tidal,fuller2012dynamical,vick2017tidal}. Even though the oscillations do not affect the angular momentum significantly, the dissipation of the excited oscillations can heat up the WD envelope, which can induce isotropic runaway nuclear fusion\,\citep{fuller2012dynamical}. The inflated envelope will be captured by the central MBH and produce a TDE flare, while the remaining core keeps inspiralling in the eccentric orbit. After a few years, the tidal stripping begins when the surviving core is also extend over the Roche lobe. 

\begin{figure*}
\gridline{\fig{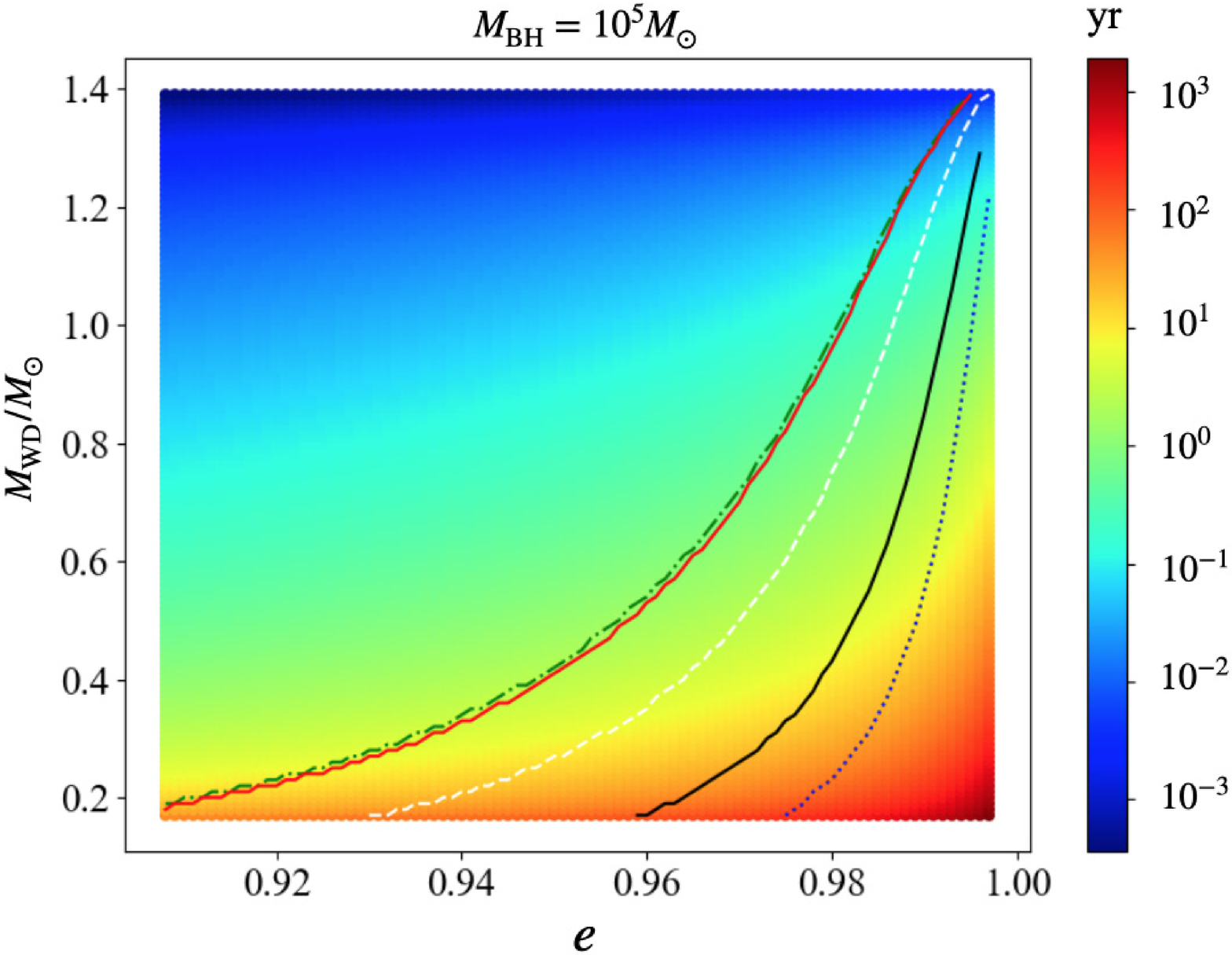}{0.5\textwidth}{(a)}
          \fig{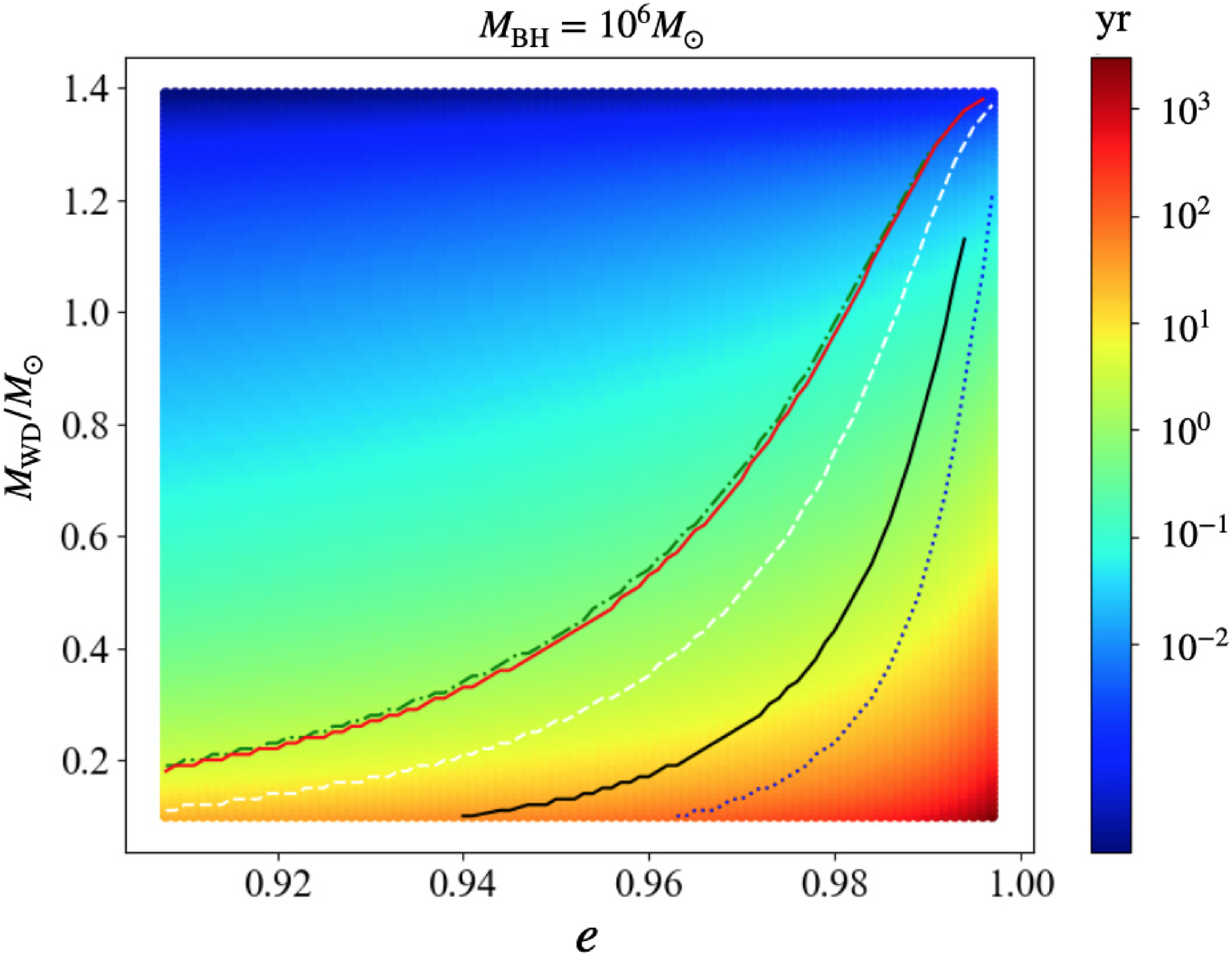}{0.5\textwidth}{(b)}
          }
\caption{The total duration time from the start of tidal stripping to fully disruption with different $e$ and $M_{\rm WD}$, where the left and right panels represent the MBH mass of $M_{\rm BH}=10^5M_\odot$ and $M_{\rm BH}=10^6M_{\odot}$ with  $\widetilde{\Omega}_s=0.5$, respectively. The solid-black, dashed-white, dotted-blue, dash-dotted-green and solid-red lines represent the relation between $M_{\rm WD}$ and $e$ of GSN 069, RX J1301.9+2747, eRO-QPE1, eRO-QPE2 and XMMSL1 J024916.6-041244, respectively. The color bar represents the total duration time in unit of year.}
\label{fig:2}
\end{figure*}

To estimate the mass loss rate in tidal stripping stage, we adopt a Roche lobe model for the mass transfer at each pericenter passage \citep[e.g.,][]{Sepinsky2007,Dosopoulou2016}. Under the quasi-static approximation, the Roche lobe radius of a non-synchronous and eccentric binary is\,\citep[][]{Sepinsky2007}
\begin{equation} \label{11}
R_L=R_p\frac{0.49q^{2/3}}{0.6q^{2/3}+{\rm ln} (1+q^{1/3})}.
\end{equation}
\noindent In our WD-MBH system, the Roche lobe radius can be well approximated by $R_L\approx 0.5 R_p q^{1/3}$\,\citep{Zalamea2010}. As the WD extends over the Roche lobe ($R_{\rm WD}>R_L$), the mass transfer will start from WD to MBH through the Lagrange point. We can obtain the initial pericenter $R_{p}\sim 2R_t$ for the tidal stripping.

The mass loss $\delta M_{\rm WD}$ in each pericenter passage can be estimated by considering the WD mass when $R_{\rm WD}>R_L$:
\begin{equation} \label{10}
    \delta M_{\rm WD}=4\pi R^2\int_0^\Delta \rho(z)dz,
\end{equation}
where $\rho(z)$ is the WD surface density given by Eq.\,(\ref{2}), and $\Delta/R_{\rm WD}\equiv (R_{\rm WD}-R_L)/R_{\rm WD}\ll 1$. The WD radius will expand as the mass decreases, therefore the tidal radius will also increase. After thousands of tidal stripping, the pericenter distance $R_p$ will be smaller than $R_t$ and finally the remaining WD will be fully tidal-disrupted. In the first $10^3 \sim 10^4$ orbits of stripping phase, where WD only loses $\sim 1\%$ of its mass, the orbital evolution is mainly regulated by gravitational radiation. In the last several hundreds of orbits the WD loses the $\sim 70\% M_{\rm WD}$ which is consistent with\,\cite{Zalamea2010}, while the remaining $\sim 30\% M_{\rm WD}$ will be fully disrupted when $R_p<R_t$. 

In Figure\,\ref{fig:2}, we present the total duration time from the tidal stripping to fully disruption with different parameters. We find that the total duration time range from hours for $M_{\rm WD}>1 M_\odot$ to several thousands years for $M_{\rm WD}<0.2 M_\odot$ for $M_{\rm BH}=10^{5-6}M_\odot$, $e=0.9-1$ and $\widetilde{\Omega}_s=0.5$. For typical parameters of $M_{\rm BH}=10^5M_\odot$, $M_{\rm WD}=0.5M_\odot$ and $e=0.98$, the total duration time is about several decades.

\subsection{The radiation of the accreted matter} \label{sec:2.4}
\begin{figure}
\centering
\includegraphics[scale=0.23]{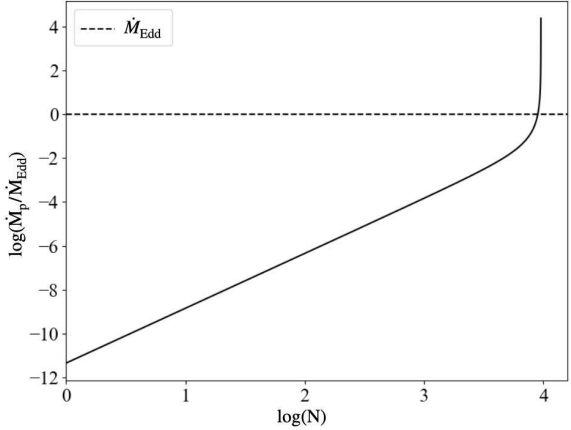}
\caption{The peak fallback rate near the pericenter radius in each orbit (solid line), where the dotted line is the Eddington accretion rate. Here, $M_{\rm BH}=10^{5}M_\odot$, $M_{\rm WD}=0.45M_\odot$, $e=0.988$ and $\widetilde{\Omega}_s=0.3$}
\label{fig:3}
\end{figure}

The disrupted matter will eventually fall into the BH with the fallback rate $\dot{M}_p$, which is empirically given by\,\citep{Rees1988,Phinney1989,Evans1989}:
\begin{equation}\label{eq:fallback}
\dot{M}_p=\delta M_{\rm WD}/3p_{\rm min},
\end{equation}
where $p_{\rm min}$ is the shortest Keplerian orbital period \citep{Ulmer1999,wang2019bright}. The peak fallback rate at pericenter along the orbital evolution is shown in Figure\,\ref{fig:3}, where the typical parameters $M_{\rm BH}=10^{5}M_\odot$, $M_{\rm WD}=0.45M_\odot$, $e=0.988$ and $\widetilde{\Omega}_s=0.3$\,\citep[$\Omega_s \sim \Omega_p$, see][]{vick2017tidal} are used. The peak accretion rate is quite low in the initial stripping stage (e.g. much less than 1\% of Eddington rate), and the radiation should too faint to be observed. Then the fallback rate will rapidly increase, when the radius of the WD increases as the WD mass decreases, which can reach $10^{5}\dot{M}_{\rm Edd}$ in last several tens orbits. 

For a typical full TDE, the fallback rate and the light curve typically follow $t^{-5/3}$\,\citep{Rees1988}. However, the fallback rate for the partial TDE of the WD envelope and the tidal stripping stage are much steeper than the full TDE, where the light curve may follow $t^{-n}$ with $n\approx 2-5$\,\citep{Ryu2020}. In this work, we adopt a power law of $n=9/4$\,\citep{Giuseppe2011,Coughlin2019} for the partial TDE of the accreted WD envelope inflated from the tidal nova and the tidal stripping, while $n=5/3$ for the fully disruption in the last passages of the remaining core to plot the typical light curve.

For the early tidal stripping stage the accretion is sub-Eddington. For the disruption of the envelope, the late stripping stage and the tidal disruption, the accretion is super-Eddington instead. Here, we adopt the slim disk model to jointly describe the super/sub Eddington accretion\,\citep{1988Slim}, since the slim disk will smoothly change into the standard thin disk at sub-Eddington accretion rate. We consider a steady accretion disk without winds for a rough estimation. The viscosity parameter $\alpha=0.1$ and the outer radius $R_{\rm circ}\sim 2R_p$\,\citep{Bonnerot2016} are used .

\begin{figure}
\centering
\includegraphics[scale=0.23]{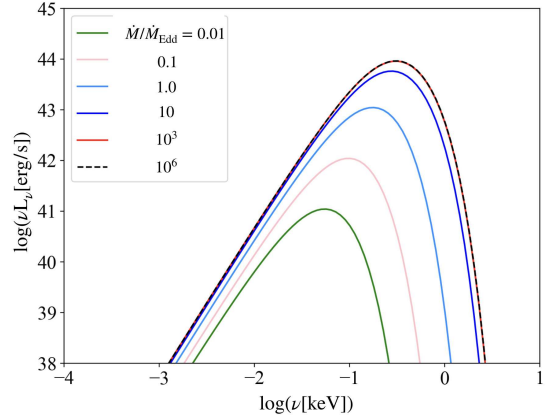}
\caption{The spectral energy distribution for the slim disk model with different accretion rates, respectively. The MBH mass is $10^5M_\odot$}.
\label{fig:4}
\end{figure}

\section{Results and discussions} \label{sec:results and discussions}
\subsection{light curve}

\begin{figure}
\centering
\includegraphics[scale=0.23]{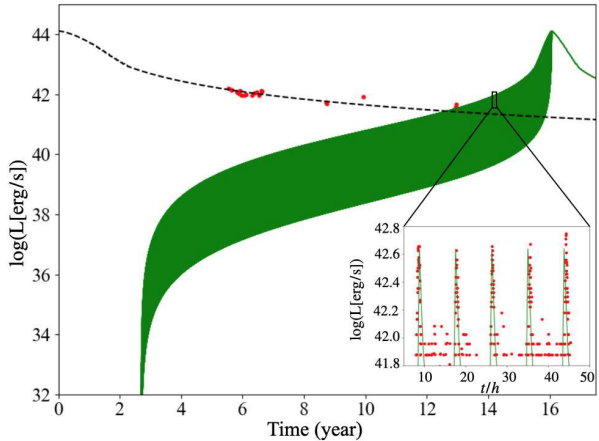}
\caption{The $0.4-2$\,keV light curve for the entire tidal interaction event in the with $M_{\rm BH}=10^5M_\odot$, $M_{\rm WD}=0.45M_\odot$, $e=0.988$ and $\widetilde{\Omega}_s=0.3$ for a typical QPE source of GSN 069. The green solid line represents the tidal stripping stage. The observational timescale ($\sim$ year) shown in the figure is too long to resolve the period ($\sim$ hour) of QPEs. So we insert a panel to show the details with higher time resolution. The dashed black line represents the disruption of WD envelope with mass $10^{-2} M_{\rm WD}$ which was ignited as tidal nova. The red dots are the observational data of TDE-like flare and QPEs of GSN 069, which are adopted from \cite{shu2018} and \cite{Miniutti2019}. } 
\label{fig:5}
\end{figure}

In Figure\,\ref{fig:4}, the spectral energy distribution (SED) is shown for the dimensionless accretion rate $\dot{M}/\dot{M}_{\rm Edd}=10^{-2} -10^{6}$ for a typical MBH mass of $10^5M_\odot$. The peak frequency of the SED stays in the soft X-ray waveband $\sim 0.1-1$ keV for the sub/super accretion rate, which is consistent with the observation of TDE-like flare and QPEs in GSN 069. The contribution from disk radiation is small in the optical/UV band compared to that in X-ray bands, which is consistent with the observation result that no periodic optical eruptions shown in these QPE sources\,\citep[][]{Miniutti2019,Arcodia2021}. In addition, the peak frequency and the luminosity of SED is roughly unchanged for $\dot{M}/\dot{M}_{\rm Edd}>10$ at a given BH mass. The saturation of the peak frequency and the bolometric luminosity is due to the fact that most of accretion energy is advected into the BH instead of radiating away because of the strong photon trapping effect in the supper Eddington accretion stage. The theoretical works also suggest that outflows/winds may be driven from the surface of the slim disk due to the radiation pressure, which will reduce the accreting material and disk luminosity\,\citep[e.g.,][]{Gu2012,Cao2015,fengjun2019}. The peak frequency of disk and the bolometric luminosity may be not affected a lot if the accretion rate is still super Eddington near the MBH horizon. However, the light curve of last several tens orbits may be affected (see Figure\,\ref{fig:5} with detailed discussion as follows) if there is an upper limit for gas swallowed by MBH\,\citep[e.g.,][]{Cao2015,fengjun2019}.

Considering the pericenter radius $R_p \sim 2R_t$ and $R_p^3/(1-e)^3=GM_{\rm BH} T^2/4\pi^2$ for an EMRI orbit ($T$ is period of the orbit), we can establish a relation between eccentricity $e$ and the WD mass $M_{\rm WD}$: $2R_{\rm WD}/(1-e)\sim (GM_{\rm WD}T^2/4\pi^2)^{1/3}$, where $R_{\rm WD}$ is given by Eq.\,(\ref{1}). We show the relation between $M_{\rm WD}$ and $e$ for five QPEs who have different recurrence time in Figure\,\ref{fig:2}. The black, white, blue, green and red lines represent GSN 069, RX J1301.9+2747, eRO-QPE1, eRO-QPE2 and XMMSL1 J024916.6-041244, respectively. The fallback rate in the tidal stripping phase is determined by the amount of the matter that stripped at each pericenter passage (e.g., Eq.\,\ref{10}). By assuming the peak luminosity of the QPEs is caused by the fallback rate of the stripping matter, it is possible to constrain the model parameters if combining with the observational period for a given mass of MBH. Observationally, the light curve for the entire tidal interaction event in the soft X-ray bands is shown in Figure\,\ref{fig:5}. For a typical BH mass of $M_{\rm BH}=10^5M_\odot$ and WD rotation rate $\widetilde{\Omega}_s=0.3$, we find that $M_{\rm WD}=0.45M_\odot$ and $e=0.988$ are possible to reproduce the $\sim$ 9-hours period and $\sim 5\times 10^{42} \rm erg/s$-peak luminosity as found in GSN 069. The black dashed line represents the tidal disruption of the possible WD envelope with mass of $10^{-2}M_{\rm WD}$ a couple of years before the QPEs, was ignited as tidal nova\,\citep[][]{fuller2012tidal,fuller2012dynamical}. The green solid line shows the light curve in the tidal stripping stage and also the fully tidal disruption in the last passages. It should be noted that the period of each burst is quite short ($\sim$hour) compare to the interested observational time scale ($\sim$year), and we show a short segment of time with a higher time resolution ($\sim$hours) in the Figure\,\ref{fig:5} to better illustrate the periodic QPE flares. The red dots are the observational data of GSN 069\,\citep[see][]{shu2018,Miniutti2019}. The peak luminosity will reach the upper limit at the last passages to the pericenter. As the TDE-like flare decays and QPE peak luminosity increases, the periodic signals will be observed at the certain stage. Observationally, \cite{sheng2021evidence} found that the GSN 069 has an abnormal C/N abundance in its UV spectrum, which supports that the TDE is happened a couple of year before the QPEs. Furthermore, our model is also supported by the observational evidence of tidal disruption events in XMMSL1 J024916.6-041244\,\citep{Joheen2021}. This TDE-QPE connection is not found in the other three sources, which may be due to the fact that the WDs have very different envelope masses. The WD with low envelope mass can not create a sufficiently bright TDE.

\begin{figure*}
\gridline{\fig{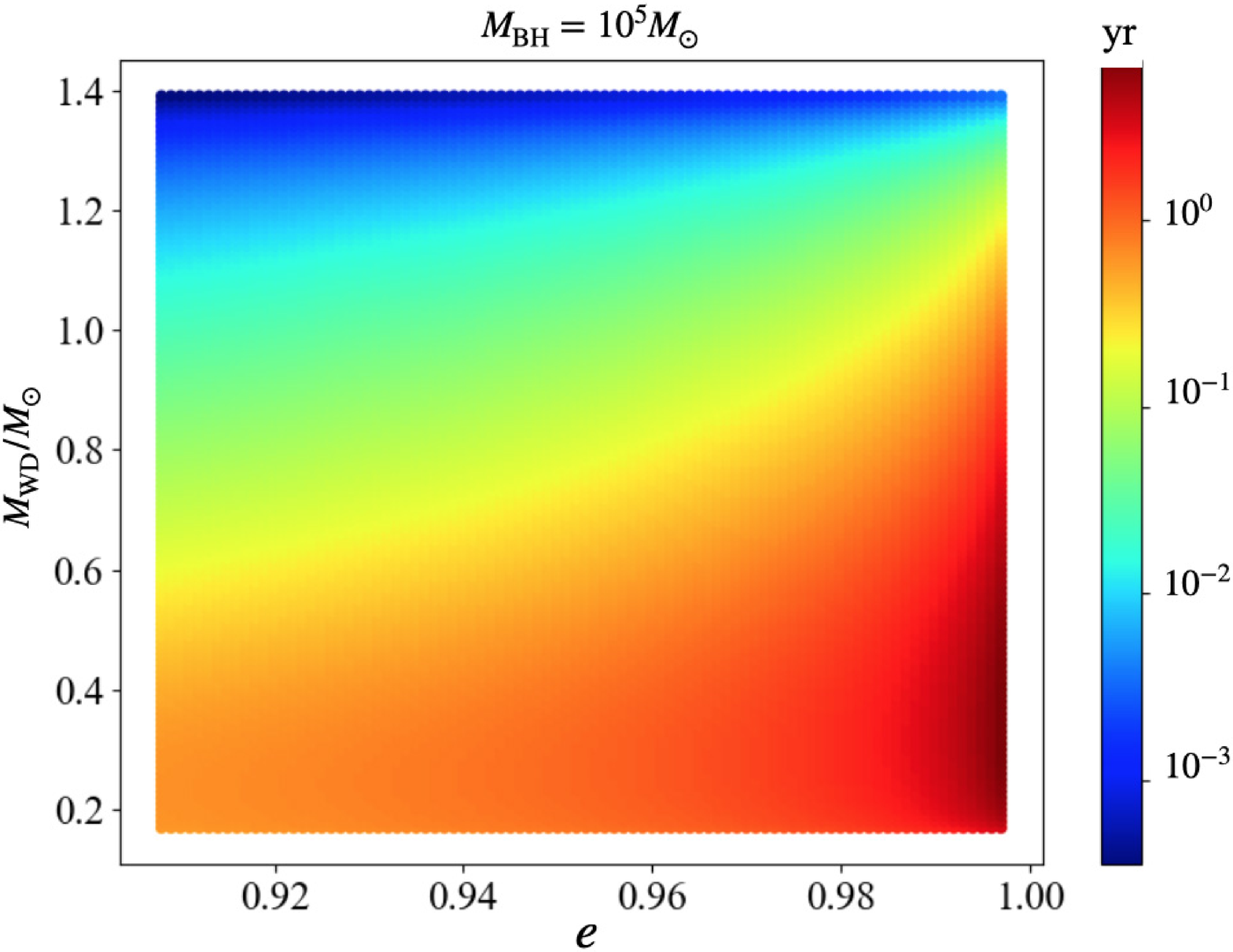}{0.5\textwidth}{(a)}
          \fig{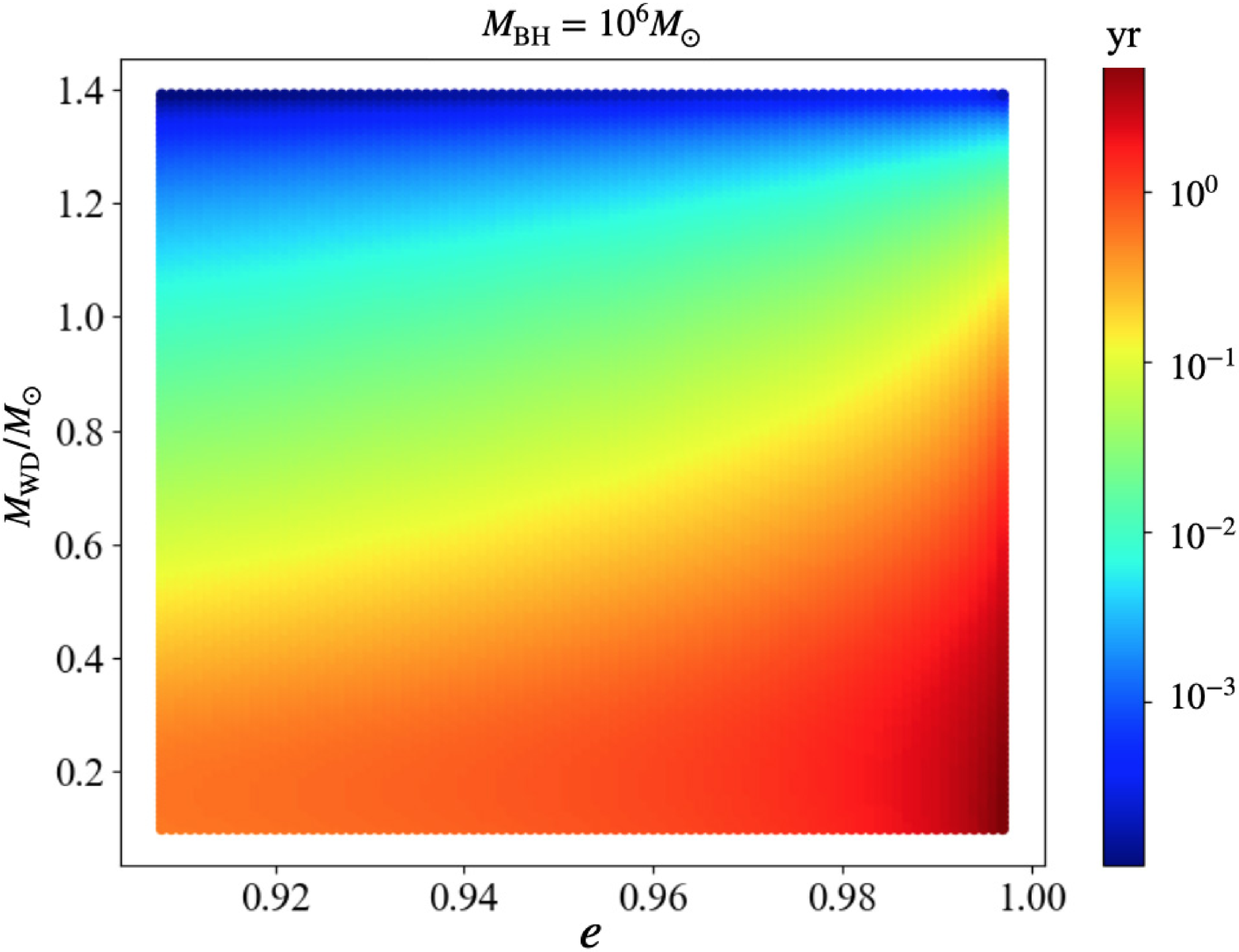}{0.5\textwidth}{(b)}
          }
\caption{The remaining time of tidal stripping event after the peak luminosity of QPEs is up to $10^{42} \rm erg/s$, where the left and right panels represent the MBH mass of $M_{\rm BH}=10^5M_\odot$ and $M_{\rm BH}=10^6M_{\odot}$ with  $\widetilde{\Omega}_s=0.5$, respectively. The color bar represents the remaining time in unit of year.
}
\label{fig:6}
\end{figure*}

\subsection{Remaining time of the QPE source} \label{sec:remainingtime}

For given BH mass, it is possible to estimate the remaining time for these QPE sources, which can be used to test this model with the following X-ray monitoring. The remaining time in our model is defined as the duration from the QPE-observed time to the fully disruption of WD. In Figure\,\ref{fig:6}, we present the remaining time after the peak luminosity is up to $10^{42} \,\rm erg/s$ of QPE sources at certain parameter spaces ($M_{\rm BH}$, $M_{\rm WD}$ and $e$). For the identified QPE sources, the results are listed as follows. For the GSN 069 with $M_{\rm BH}\sim 4\times 10^5\,M_\odot$ and $L\sim 5\times 10^{42} \,\rm erg/s$, we find the remaining time is $\sim 1.2 \,\rm years$. For RX J1301.9+2747 with $M_{\rm BH}\sim 1.8\times 10^6\,M_\odot$ and $L\sim 1.4\times 10^{42} \,\rm erg/s$, the remaining time is $\sim 1 \,\rm year$. For eRO-QPE1 with $M_{\rm BH}\sim 9.1\times 10^5\,M_\odot$ and $L\sim 3.3\times 10^{42} \,\rm erg/s$, the remaining time is $\sim 1.5\,\rm years$. For eRO-QPE2 with $M_{\rm BH}\sim 2.3\times10^5\,M_\odot$ and $L\sim 1.0\times 10^{42} \,\rm erg/s$, the remaining time is $\sim 1\ \rm year$. For XMMSL1 J024916.6-041244 with $M_{\rm BH}\sim 8.5\times 10^4\,M_\odot$ and $L\sim 3.4\times 10^{41} \,\rm erg/s$, the remaining time is $\sim 2 \,\rm years$. It should be noted that we calculated the remaining time by assuming a fast-spinning WD ($\widetilde{\Omega}_s=0.5$, see Figure\,\ref{fig:6}). If the WD spins slowly, dynamical tidal effects will dominate orbital evolution and cause tidal stripping process to slow down (see Eq.\,\ref{Rp}), which could increase the total duration time and the remaining time of QPEs.

\subsection{The issue of event rate}
In \cite{metzger2021}, an estimation of the event rate $\leq 10^{-10} \rm yr^{-1} \rm gal^{-1}$ is given for the highly eccentric binary systems formed by the Hills mechanism. However, due to the possible complicated environment of the galaxy center, we think it could still be possible that the event rate maybe much higher. The event rate $\dot{N}_{\rm QPE}=\dot{N}_{\rm Hills} f_{\rm WD} f_{\rm T}$ is mainly contributed by the rate of Hills mechanism itself $\dot{N}_{\rm Hills}$, the fraction of the WD binaries (to the total number of binary stellar objects) near the galaxy center $f_{\rm WD}$, and the fraction of the WD binaries $f_{\rm T}$ with certain orbital distances which can produce 10 hours orbital period\,\citep[see][for details]{metzger2021}.  

Firstly, the rate of Hills mechanism in the core galaxies is $\dot{N}_{\rm Hills}\sim 10^{-5}-10^{-3} \rm yr^{-1} \rm gal^{-1}$\,\citep[e.g.,][]{Yu2003}. However, it will be much higher ($\dot{N}_{\rm Hills}\sim 10^{-2}-10^{-1} \rm yr^{-1} \rm gal^{-1}$) in the cusp galaxies\,\citep[e.g.,][]{Yu2003,Wang2004,Mastrobuono2014,Hugo2020}. The observational data indicates that only 10 percent of the galaxies are cusp ($p_{\rm cusp}=10\%$)\,\citep[see][]{Nicole2019,Shiyong2021}, therefore the Hill mechanism rate should be estimated as $\dot{N}_{\rm Hills}=\dot{N}^{\rm cusp}_{\rm Hills}p_{\rm cusp}+\dot{N}^{\rm core}_{\rm Hills}p_{\rm core}\sim 10^{-3}-10^{-2}  \rm yr^{-1} \rm gal^{-1}$ where $p_{\rm core}=1-p_{\rm cusp}$.

Secondly, the $f_{\rm WD}$ which is the fraction of the WD binaries near the galaxy center could be a few times larger than the double WDs' result ($\sim 0.002$) estimated in \cite{metzger2021}. This could happen because of the supernova feedback process near the galaxy center, which is also a potential mechanism for the evolution of cusp galaxy structure to the core galaxy structure\,\citep[e.g.,][]{Gelato1999,deBlok2010,Burger2021}. The supernova process will drive the interstellar gas from the galaxy center to the outside, while the COs such as WDs and neutron stars are still left there. This will suppress the star formation near the galaxy center and thereby increase the fraction of the WD binaries, which means that this fraction may not be homogeneously distributed over the galaxy radius. It is reasonable to assume that the fraction is a few times higher than $0.002$.

Thirdly, the fraction of the double WDs with orbital distance $a_b$ less than the critical value ($\sim 0.1R_\odot$) is probably underestimated ($f_{\rm T} \sim T_{\rm bin}/T_{\rm Hubble}\sim 4\times 10^{-6}$, where $T_{\rm bin}$ is the lifetime of WD binary and $T_{\rm Hubble}$ is the Hubble time) in \cite{metzger2021}. However, \cite{metzger2021} estimated the lifetime of WD binary by only considering gravitational radiation. Although the dynamical tides can hardly effect the orbital evolution\,\citep[][]{fuller2012tidal,fuller2012dynamical}, the lifetime of WD binaries can be increased by 1-2 orders of magnitude if considering the mass transfer between WD binaries compared that just considering gravitational radiation. Furthermore, it is possible that WDs may not be tidal stripped immediately after being captured. For example, the critical distance of GSN 069 can be up to $\sim 0.2R_\odot$ since there is an about ten-years delay between the separation of the WD binary and the start of nine-hours period QPEs. Moreover, recent cosmological numerical simulations has been done for predicting the LISA double WD population by \cite{Lamberts2019}. They showed that the fraction of double WDs entering the orbital frequency $>10^{-3}$\,Hz (this frequency corresponds to the orbital distance $\sim 0.2R_\odot$) can be $10^{-3}-10^{-2}$, which is almost three orders of magnitude higher than that given in \cite{metzger2021}. The reason could be that the estimation given in \cite{metzger2021} is based on the assumption that the population of the double WDs is homogeneously distributed with respect to their age, while the cosmological simulation showed that this distribution is not homogeneous\,\citep[for details, see][]{Lamberts2019}.

Considering the above three factors, the event rate could be around $\dot{N}_{\rm QPE}\sim 10^{-8}-10^{-6} {\rm yr}^{-1}{\rm gal}^{-1}$. The observed number of QPEs by eROSITA can be estimated by $N_{\rm QPE}=\dot{N}_{\rm QPE}N_{\rm gal}\tau$, where $N_{\rm gal}$ is the total number of galaxies that will be surveyed by eROSITA by the end of 2023 and $\tau$ is the lifetime of QPEs. The comoving volume within the redshift $z=0.05$ of the most distant QPE event (eRO-QPE1) is $V\sim 0.04 \rm Gpc^3$. Using the local density $\sim 5\times 10^7 \rm Gpc^{-3}$ of dwarfs galaxies with total stellar mass $\sim 10^9M_\odot$\,\citep{Paul2015}, the number of local galaxies is $N_{\rm gal}\sim 2\times 10^6$ within the redshift $z=0.05$. We find that $N_{\rm QPE}\sim 1-20$ if adopting the typical active lifetime of QPEs is $\sim 10$ years (see Figure\,\ref{fig:2}), which is roughly consistent with the predicted 10-15 eROSITA-discovered QPEs by the end of 2023\,\citep[][]{Arcodia2021}. The X-ray monitoring based on eROSITA and future Einstein Probe (EP) can further shed light on event rate of QPEs and the physics of the tidal disruption of WD\,\citep{Yuan2016,yuan2018}.

\begin{figure}
\centering
\includegraphics[scale=0.27]{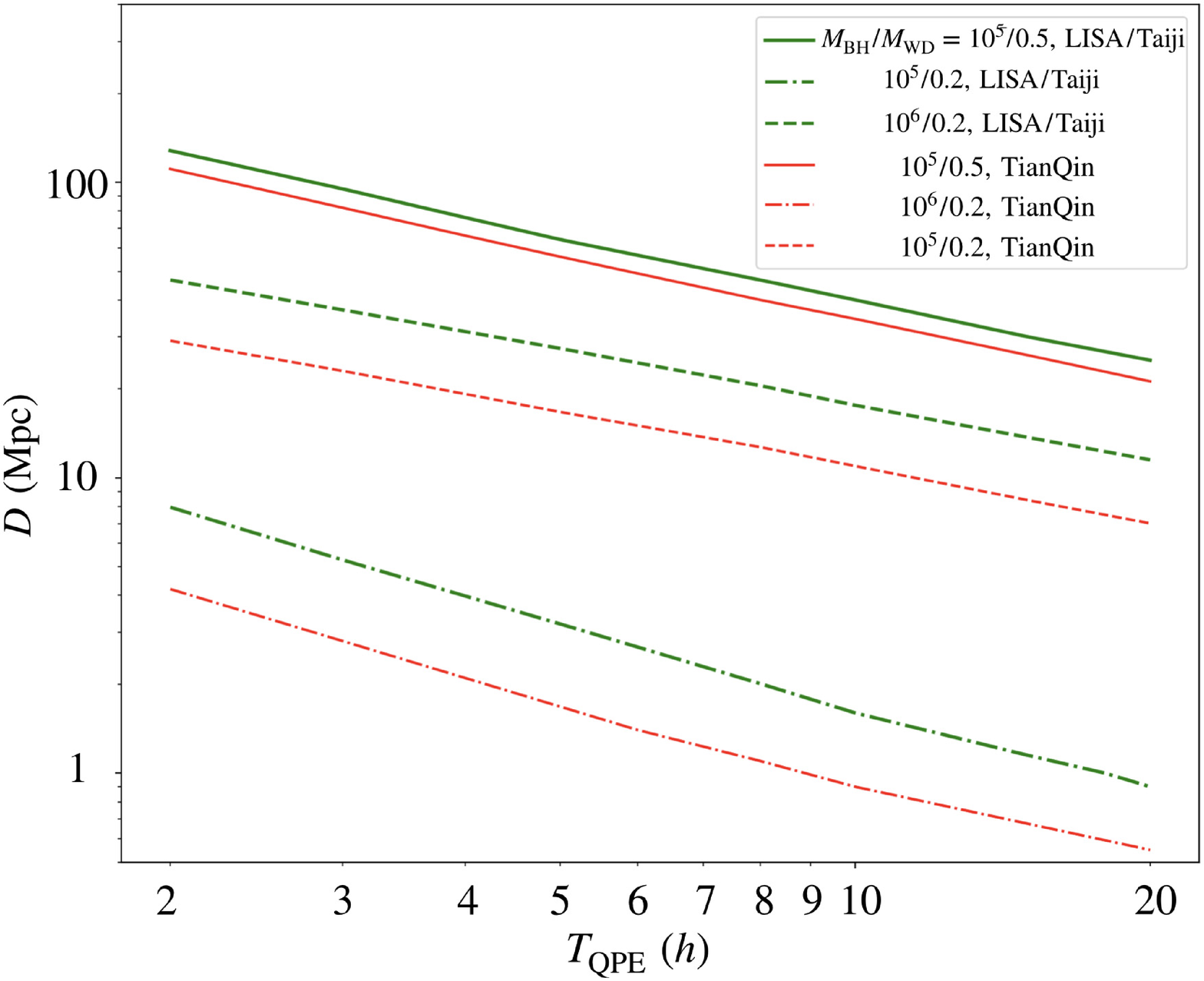}
\caption{ The relation between the distance limit $D$ of GW detection (SNR>5) and the period $T_{\rm QPE}$ in typical $M_{\rm BH}$ and $M_{\rm WD}$. The green and red lines represent the detection limit of LISA/Taiji and TianQin, respectively }
\label{fig:7}
\end{figure}

\subsection{The possibility for space-borne GW detection} \label{sec:remainingtime}

EMRIs are important targets for future space-bore detectors\,\citep[e.g.,][]{Pau2007,HanWenbiao2018,Chenxian2018,Chenxian2021}
such as LISA, TianQin and Taiji\,\citep[e.g.,][]{Luo2015TianQin,Travis2019}. 
In a recent work, \cite{Chenxian2021} estimated the effective strain for these QPEs sources, and they found that the GW from these several QPE sources cannot be detected by above proposed space-born GW detectors. Here, we simply estimate the possible detection distance for the EMRIs as expected in QPE sources. The MBH-WD EMRI in our model has a highly eccentric orbit so that the GW radiation is spread into a wide range of harmonic frequencies\,\citep[e.g.,][]{Peters1963,finn2000gravitational,Barack2004}. The characteristic strain $h_{c,n}$ of the $n$-th harmonic component, which corresponds to the frequency $n/T$\,\citep[$T$ is the orbital period,][]{finn2000gravitational,Barack2004,Sesana2020,Chenxian2021},
\begin{equation}
    h_{c,n}(f)=\frac{1}{\pi D}\sqrt{\frac{2G\dot{E}_n}{c^3\dot{f}_n}},
\end{equation}
where $D$ is the distance from the QPE source to the detector, $\dot{E}_n$ is the $n$-th harmonic gravitational wave power radiated by the WD-MBH binary and $\dot{f}_n$ describes the time evolution of the $n$-th harmonic GW frequency. In our model, the GW radiation timescale is many orders of magnitude longer than the mission duration of future space-bore detectors, e.g., $t_{\rm LISA}=4\ \rm years$. In this case, the effective strain, which is directly proportional to the signal-to-noise ratio (SNR), is $h_{{\rm eff},n}=h_{c,n}\sqrt{\Delta f/f}$ where $\Delta f$ is the increment of frequency during the observational period\,\citep[][]{Barack2004,Sesana2020,Chenxian2021}. The total effective strain $h_{\rm eff}$ can be calculated with 
\begin{equation}\label{13}
    h^2_{\rm eff}(f)=\sum_{n=1}^{\infty}h^2_{{\rm eff},n}
\end{equation}
The corresponding SNR can be calculated with 
\begin{equation}\label{14}
    {\rm SNR}^2=\int  \frac{h^2_{\rm eff}}{fS_{\rm n}(f)} {\rm d\ \rm ln}\ f,
\end{equation}
where $S_{\rm n}(f)$ is the total (sky-average) LISA, TianQin or Taiji sensitivity. 

Assuming the 4-years observational time, we evolve the orbital parameters from $t=0$ (when the tidal stripping start) to $t=t_{\rm LISA}$ to obtain the orbital frequency within the range $f_{\rm orb,min}<f_{\rm orb}<f_{\rm orb,max}$. We present the relation between the initial period $T_{\rm QPE}$ of QPEs and the threshold distance $D$ from the source to the detector for the threshold SNR to be $5$ in Figure\,\ref{fig:7}, where the thick and thin black lines represent the threshold detection distance of LISA/Taiji and TianQin, respectively. For a typical MBH mass of $10^5\msun$ and WD mass of 0.5$\msun$, the threshold detection distance is only $\sim 20-100$ Mpc for the QPE with 2-20\,hours period. The threshold  detection distance will be closer for about one order of magnitude for the case $M_{\rm BH}=10^5\msun$ and $M_{\rm WD}=0.2 \msun$ at given QPE period.

\section{Summary} \label{sec:summary}

In this work, we provide a model to explain the possible connections TDE and QPE phenomena as observed in GSN 069. The WD binary is splitted by the tidal force of MBH, and one WD component can be captured by the MBH while its companion will be ejected away. This Hills mechanism can naturally form a highly eccentric orbit, with an event rate $10^{-3}-10^{-2}{\rm yr}^{-1}{\rm gal}^{-1}$. The possible envelope of WD can be heated and induce a runaway fusion due to the internal oscillations of WD by the tidal force. The inflated envelope captured by MBH will produce a TDE. After a couple of years, the tidal disruption of the remaining WD will form the QPEs when the WD core extends over the Roche lobe. Our model can well explain the TDE and QPE features as observed in GSN 069. The TDE may be absent or weak if the WD has no or weak envelope. We constrain the model with the period and peak luminosity of five QPEs, and we find the remaining time for these WD-MBH systems are only a couple of years, which can be used to test the proposed model. 

\begin{acknowledgments}
We appreciate Dong Lai, Huan Yang, Ning Jiang, Weihua Lei, Yan Wang and Xinwen Shu for many useful discussions. We also thank our anonymous referees for their important and insightful comments on our manuscript. Y.M. is supported by the university start-up fund provided by Huazhong University of Science and Technology. Q.W. is supported by the NSFC (grants U1931203), the science research grants from the China Manned Space Project (No. CMS-CSST-2021-A06) and the National Key Research and Development Program of China (No. 2020YFC2201400).  
\end{acknowledgments}

\appendix

\section{Hills mechanism} \label{appendix}
In this appendix, we provide a derivation of the orbital eccentricity of the bounded WD-MBH system generated by Hills mechanism following\,\citep{Pau2018,Pau2020}.

Hills mechanism provides a physical scenario to produce a highly eccentric WD-MBH binary system through a 3-body exchange between a WD binary and a MBH \citep[e.g.,][]{Hills1988,Bromley2012,Brown2015}. When the WD binary passes closed to a MBH, the tidal force can make the binary split. The split radius $R_{\rm split}$ can be estimated using the fact that the split happens when the gravitational tidal force becomes comparable to the gravitational force that bounds the binary, i.e.
\begin{equation}
\frac{GM_b}{a_b^2}\sim \frac{GM_{\rm BH}a_b}{R_{\rm split}^3},
\end{equation}
and the result is: $R_{\rm split}\sim a_b (M_{\rm BH}/M_{b})^{1/3}$, where $M_{\rm BH}$ is the mass of MBH, $a_b$ and $M_b$ is the separation and total mass of the WD binary. After the splitting, \cite{Hills1988} suggests that one binary component becomes bounded to the MBH, while the other is ejected to be a Hypervelocity star. Denote the orbital velocity of the binary as $v_{\rm orb}\sim \sqrt{GM_b/a_b}$ and the velocity of the centre-of-mass as $v_{\rm com}\sim \sqrt{GM_{\rm BH}/R_{\rm split}}$, energy conservation leads to the following relation among the the semi-major axis of the orbit of the captured WD $a_{\rm cap}$, $v_{\rm com}$ and $v_{\rm orb}$:  
\begin{equation}
    \frac{1}{2}(v_{\rm com}-v_{\rm orb})^2-\frac{1}{2}v_{\rm com}^2=-\frac{GM_{\rm BH}}{2a_{\rm cap}}.
\end{equation}
Since $v_{\rm com}\gg v_{\rm orb}$, the $a_{\rm cap}$ can be estimated as 
\begin{equation}
a_{\rm cap}\sim a_b(M_{\rm BH}/M_b)^{2/3},
\end{equation}
using the above formula\,\citep{Pau2018,Pau2020}. The triggering of the WD tidal stripping requires that the pericenter distance of the bound WD satisfies $R_p\sim 2R_t$\,\citep{Zalamea2010}, where $R_t=R_{\rm WD}(M_{\rm BH}/M_{\rm WD})^{1/3}$ is the tidal radius for the tidal stripping of the WD by the MBH. Thus, the eccentricity of the bound WD $e_0=1-R_p/a_{\rm cap}$ can be given by 
\begin{equation}
e_0=1-2R_t/a_{\rm cap}\sim1-R_{p}/R_{\rm split}(M_{\rm BH}/M_b)^{-1/3},
\end{equation} 
where we have used the relations of $R_{\rm split}\sim a_b (M_{\rm BH}/M_{b})^{1/3}$, $R_t\sim R_{\rm WD}(M_{\rm BH}/M_{\rm WD})^{1/3}$ and $a_{\rm cap}\sim a_b(M_{\rm BH}/M_b)^{2/3}$.

\bibliography{ref}{}
\bibliographystyle{aasjournal}

\end{document}